\documentstyle[aps,pre,epsf]{revtex}

\author{Rav\'a da Silveira and Mehran Kardar\\
Department of Physics\\
Massachusetts Institute of Technology\\
Cambridge, Massachusetts 02139}
\title{Critical Hysteresis for $n$-Component Magnets
}
\date{\today
}

\begin{document}

\maketitle
\begin{abstract}
Earlier work on dynamical critical phenomena in the context of magnetic
hysteresis for uniaxial (scalar) spins, is extended to the case of a
multicomponent (vector) field. From symmetry arguments and a perturbative
renormalization group approach (in the path integral formalism), it is found
that the generic behavior at long time and length scales is described by the
scalar fixed point (reached for a given value of the magnetic field and of
the quenched disorder), with the corresponding Ising-like exponents. By
tuning an additional parameter, however, a fully rotationally invariant
fixed point can be reached, at which all components become critical
simultaneously, with ${\cal O}\left( n\right) $-like exponents. Furthermore,
the possibility of a spontaneous non-equilibrium transverse ordering,
controlled by a distinct fixed point, is unveiled and the associated
exponents calculated. In addition to these central results, a didactic
``derivation'' of the equations of motion for the spin field are given, the
scalar model is revisited and treated in a more direct fashion, and some
issues pertaining to time dependences and the problem of multiple solutions
within the path integral formalism are clarified.
\end{abstract}

\section{ Introduction}

In a great variety of non-equilibrium situations, critical behavior is
observed as a system evolves from one of its possible states to another.
Some examples are charge density waves, fluctuating interfaces and lines,
cracks and fractures, and the Barkhausen effect in magnets. These systems
evolve, respectively, from a state without current to a state with current,
from a stationary to a moving state, from a connected to a ruptured state,
from a downward to an upward magnetization. Under specific conditions 
({\it  i.e.} preparation of the system), the transition between the two states is
critical (or continuous), exhibiting diverging correlation lengths, and
scaling laws. The qualitative descriptions of the dynamics of the different
physical situations mentioned are very similar in the key parameters and
mechanisms which govern criticality. The quantitative descriptions are also
close, in that the dynamics of motion can be described by continuum field
equations, and share many common features. In this article we focus on
magnetic systems. More specifically, on a lattice of spins with
ferromagnetic exchange (coupling). While our qualitative and quantitative
analyses will be in this framework, some aspects of the discussion may apply
to other systems, such as depinning transition of flux lines or fractures in
disordered media.

What do we mean by ``the key parameters and mechanisms which govern
criticality''? Consider a ferromagnet in an external magnetic field which
increases slowly from $-\infty $ to $+\infty $. Each spin feels a local
field equal to the average of the surrounding spins multiplied by the
coupling constant ($JM_i$ in the case of the $i^{th}$ spin), plus the
external magnetic field $H$. At $H=-\infty $,  all the spins
point ``downward'' ($M_i=-1$ for unit spins), thus $JM_i=-J$ initially. At
zero temperature, each spin simply points in the direction of the local
field $JM_i+H$, and so none of the spins change before $H$ reaches $J$, at
which point they all flip upwards. The magnetization $M$ thus jumps from $-1$
to $+1$ at $H=J$. This  scenario for a perfectly clean system is modified
by introducing some disorder. At each lattice point occupied by a spin,
add a random field, $h_i$, to the local field, $JM_i+H$. (Specifically,
let $h_i$ be an uncorrelated random variable, chosen from a Gaussian
distribution centered at zero.) Then, the spins flip in a much less coherent
way: as soon as $JM_i+H+h_{i\text{ }}$becomes positive, the $i^{th}$ spin
flips. The upward $h_i$'s enhance the increase in magnetization for low $H$,
whereas the downward $h_i$'s suppress it for high $H$. This results in the
reduction of the magnitude of the jump in magnetization. Clearly, if we
broaden the random field distribution, {\it i.e.} increase the amount of
disorder, the discontinuity in $M$ is further suppressed, until the curve $M(H)$
eventually becomes smooth for a high enough disorder (Fig.~1). We can
imagine a sequence of hysteresis curves, corresponding to a succession of
increasing amounts of disorder, say for the variance of the random field, 
$\overline{h^2}$, going from zero to infinity. The curve displays a
discontinuity for small $\overline{h^2}$, and is smooth for large 
$\overline{h^2}$. The transition between the two regimes occurs at a critical point
reminiscent of continuous or second-order phase transitions 
\cite{dahmen,prb,thesis,leshouches}: at the critical amount of disorder the
discontinuity collapses to a point, at which the slope is infinite. This is
referred to as a {\it critical hysteresis}, for which, at a given magnetic
field, the susceptibility diverges. The amount of disorder and the magnetic
field are the two parameters we have to tune to observe criticality.

\begin{figure} 
\epsfxsize=16truecm 
\vspace*{.6truecm} 
\centerline{
\epsfbox{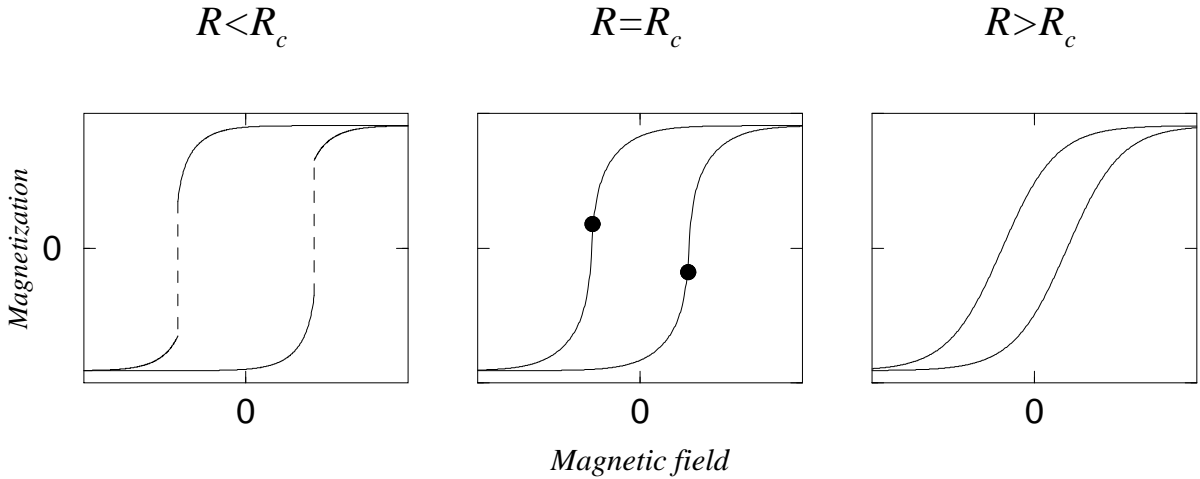} 
} 
\vspace*{.3truecm}

\caption{Schematic hysteresis curves for different values of the disorder
$R$. Left: $R<R_c$ (discontinuous hysteresis); center: $R=R_c$ (critical
hysteresis); right: $R>R_c$ (smooth hysteresis). In each case, the lower
(upper) curve corresponds to an increasing (decreasing) magnetic field.}
\vspace*{.6truecm}
\end{figure}

In the last few years, disorder-induced critical hysteresis in magnets has
been the subject of much interest\cite
{dahmen,prb,thesis,leshouches,narayan,zapperi}. Dahmen, Sethna, and others
studied{\bf \ }this problem via a mean-field approximation, one-loop
momentum-space renormalization, and numerical simulations. Also, they
describe a mapping of this non-equilibrium problem onto the equilibrium
random field Ising model, which can in turn be mapped (close to the upper
critical dimension) onto the pure Ising model in two lower dimensions\cite
{imry}. Throughout their work, they consider a scalar order parameter. They
study the dynamics of an Ising (or discrete) spin field driven by an
increasing magnetic field and in the presence of a random field, at zero
temperature.

The question we ask here is the following: How are the phase diagram and
critical behavior modified if the order parameter is vectorial instead of
scalar? More precisely stated: in the renormalization group (RG) framework,
is the fixed point which controls the above-mentioned hysteretic
criticality, one and the same for both the Ising and the vectorial cases?
And if not, how do the exponents differ? In equilibrium, the disordering of
scalar and vector systems is described by distinct universality classes\cite
{imry}. Also, in the closely related context of depinning transitions, the
distinction between interfaces (scalar) and flux lines (vector) was noted in
Ref.~\cite{ertas}.

The answer can readily be guessed on symmetry grounds. Indeed, symmetry
considerations lead to two distinct cases. In the first, {\it and generic one}, 
the critical hysteresis curve is such that the susceptibility diverges at
a non-vanishing value of either the magnetic field or the magnetization.
Then, although we consider continuous spins, the full rotational symmetry of
the problem is broken at the critical point, a unique preferred direction is
picked, and Ising-like critical behavior results. It is similarly argued,
in Ref.~\cite{prb}, appendix E, and Ref.~\cite{thesis}, appendix L, that
the universality class of the random field scalar model extends also to
random bonds scalar models (with a positive non-zero mean of the bonds'
values) and to random anisotropies ${\cal O}\left( n\right) $ models.
On the other hand, if both
the magnetic field and the magnetization vanish when the susceptibility
diverges, we may have a fully rotation invariant system at the critical point.
In that case, we expect ${\cal O}\left( n\right) $-like criticality, with
exponents that differ from those of the Ising model. 
Evidently, a vanishing magnetization at $H=0$ is not a sufficient
condition for a full rotational invariance. In particular, due to its
history, the system might well display higher order anisotropies, such
as, {\it e.g.} $\overline{s_{\Vert }^2} \neq \overline{s_{\bot\alpha }^2}$,
where $s_{\Vert }$ is the component of the spin field parallel to
${\bf H}$ and $s_{\bot\alpha }$ ia any parpendicular component. This
issue is resolved by a renormalization group analysis, which confirms
our various guesses. Furthermore, it discloses the possibility
of a ``transverse critical point'', corresponding to an
instability of the magnetization component {\it perpendicular} to the
external magnetic field.

The present paper is organized as follows. In Sec.~II, we construct the
equation of motion of a vector spin field. A path integral formalism is
described in Sec.~III, with which the problem of renormalizing the
equation of motion is recast into that of renormalizing a partition
function. The time dependences, as well as the subtleties associated with
``many energy minima'', are also examined. In Sec.~IV, the
renormalization group treatment of the problem is presented. First, we
define the coordinates' and fields' rescalings, and calculate the free
propagator. Then, we discuss successively the scalar and vector models. For
the latter, the different cases (hysteretic or non-hysteretic, longitudinal
or transverse criticality) are analysed, and the corresponding recursion
relations and exponents are obtained.

\section{Equations of motion}

The equations of motion for a scalar field are discussed in
Refs.~\cite{dahmen,prb,thesis}, and their generalization to a
multicomponent field is straightforward. Nonetheless, for completeness
and to emphasize our perspective, we present here a didactic introduction 
to the equations of motion for vectorial spins, 
at zero temperature\cite{thermal}.
Consider a $d$-dimensional lattice, with a spin ${\bf s}_i\in \Re ^n$ at
each site $i$, subject to a magnetic field which changes slowly from 
$-\infty $ to $+\infty $, say linearly in time, ${\bf H=\Omega }t$. The rate 
${\bf \Omega }$ can be made arbitrarily small in magnitude and points along
the first axis of our coordinates, {\it i.e.} $H_1=\Omega t\equiv H$, 
$H_2=\cdots =H_n=0$. The time dependent magnetic field implies a time
dependent energy function ${\cal H}\left( t\right) $. At zero temperature 
($T=0$), the spins simply follow the local minimum of this energy function
according to 
\begin{equation}
\eta \partial _t{\bf s}_i=-\frac{\delta {\cal H}}{\delta {\bf s}_i},
\end{equation}
starting from a uniform downward pointing configuration at $t=-\infty $. The
parameter $\eta $ controls the relaxation rate of spins towards the
time-dependent local energy minimum of ${\cal H}$. The smaller $\eta $, the
faster spins relax, and the less they lag behind the energy minimum \cite
{precession}.

The following ``key features'' guide us in constructing the Hamiltonian 
${\cal H}$. First, to describe a ferromagnet, we include in ${\cal H}$
couplings $J_{ij}$ which tend to align the spins. In addition to the
external uniform field ${\bf H}$ which drives the system, we include
quenched random fields ${\bf h}_i$. The ${\bf h}_i$'s are uncorrelated
Gaussian random variables, chosen from the distribution 
\begin{equation}
\label{distribution}\rho \left[ {\bf h}\right] =N\exp \left( -\sum_i
\frac{{\bf h}_i^2}{2R}\right) ,
\end{equation}
where $N$ is a normalization factor. For calculational convenience we shall
work with soft spins (whose magnitude can take any real value), which can be
thought of as a coarse-grained picture of a hard spin field. To avoid the
unphysical instability of spins diverging in magnitude, we introduce an
on-site potential $V\left( {\bf s}_i\right) $, which constrains the
magnitude to remain close to 1 (or some finite number). The potential $V$ is
spherically symmetric (a ``double well'' in the scalar model and a ``Mexican
hat'' in the vector model) and is expressed through its Taylor expansion
about the origin, as 
\begin{equation}
V\left( {\bf s}_i\right) =-\frac {c_1}2\,{\bf s}_i^2-\frac {c_2}4\left( {\bf s}
_i^2\right) ^2+\cdots \text{.}
\end{equation}
Whether or not $V$ is analytic at the origin is unimportant, since $\left| 
{\bf s}_i\right| $ is constrained to be close to 1 (and not to 0). The full
Hamiltonian is now given by 
\begin{equation}
{\cal H}=-\frac 12\sum_{i,j}J_{ij}{\bf s}_i\cdot {\bf s}_j+\sum_i
\left[ -{\bf H\cdot s}_i-{\bf h}_i\cdot {\bf s}_i+V\left( {\bf s}_i\right) \right] .
\end{equation}

The gradient descent with this Hamiltonian leads to the equation of motion 
\begin{equation}
\eta \partial _t{\bf s}_i=\sum_jJ_{ij}{\bf s}_j+{\bf H+h}_i-\frac{\partial V
}{\partial {\bf s}_i}\text{.}
\end{equation}
Assuming that $J_{ij}$ is a function of the separation between spins
results, in the continuum limit, in 
\begin{equation}
\eta \partial _t{\bf s}\left( {\bf x}\right) =\int d^dx^{\prime }J\left( 
{\bf x-x}^{\prime }\right) {\bf s}\left( {\bf x^{\prime }}\right) +{\bf H}
\left( t\right) {\bf +h\left( x\right) -}\frac{\partial V}{\partial {\bf s}}
\text{.}
\end{equation}
While rewriting the problem in the continuum limit, we must impose some
limit on how fine-grained the spin field ${\bf s\left( x\right) }$ may be
because of its lattice origin. In other words, ${\bf s\left( x\right) }$ is
a superposition of Fourier components whose wave numbers are restricted from
zero to some cutoff $\Lambda $.

Finally, if the exchange function decays fast enough (as its argument
increases) for its Fourier component to be non-singular at the origin of
momentum space, {\it i.e.} if $\widetilde{J}\left( {\bf q}\right)
=1-Kq^2+\cdots $, then 
\begin{equation}
\label{short-range}\int d^dx^{\prime }J\left( {\bf x-x}^{\prime }\right) 
{\bf s}\left( {\bf x^{\prime }}\right) \approx {\bf s\left( x\right) }
+K\nabla ^2{\bf s\left( x\right) },
\end{equation}
and the equation of motion can be written as 
\begin{equation}
\label{motion}\eta \partial _t{\bf s}=K\nabla ^2{\bf s}+{\bf H}+{\bf h}-
\frac{\partial \tilde V}{\partial {\bf s}},
\end{equation}
where the coefficient $c_1$ in the the expansion of the potential has been
modified in order to take the ${\bf s}\left( {\bf x}\right) $ term of 
Eq.~(\ref{short-range}) into account. 

\section{Path Integral Formalism}

In order to identify the critical properties of our model, we should ideally
solve its equation of motion. In practice, we study the behavior of 
Eq.~(\ref{motion}) under a coarse-graining transformation. This allows us to locate
and characterize a scale-invariant point. As a first step, we recast the
equation of motion into a path integral (or generating functional), which
incorporates the whole history of the system. The generating functional is
written as the sum over all paths of the exponential of some action, which
is then renormalized perturbatively. The advantage of reformulating the
problem in this way is that we can express the perturbative treatment in a
diagrammatic fashion similar to other field theories.

We define the generating functional \cite{msr} simply as the sum over all
paths of a delta function which makes each spin follow the time evolution
given by Eq.~(\ref{motion}), {\it i.e.}

\begin{eqnarray}
\label{delta} Z & = & \int
{\cal D}s\,\delta \left( {\bf s}-\text{solution of Eq.~(\ref{motion})}\right)
\nonumber \\
& = & \int {\cal D}s\,\delta \left( -\eta \partial _t{\bf s}+K\nabla ^2{\bf
s}+{\bf H}+{\bf h}-\frac{\partial \tilde V}{\partial {\bf s}}\right)\times
\left(
\text{Jacobian}\right) \text{,}
\end{eqnarray}
where ${\cal D}s$ stands for $\prod \left\{ \text{over ``all }t\text{'',
``all }{\bf x}\text{'', }\alpha =1,\cdots ,n\right\} ds_\alpha \left( {\bf x}
,t\right) $. The Jacobian merely normalizes the value of $Z$ to unity, and
we shall henceforth ignore it \cite{jacobian}. Let us rewrite the delta
function in its representation as the integral of an exponential [$2\pi
\delta \left( f\right) =\int d\hat s\exp \left( i\hat sf\right) $]. Then,
after absorbing a factor $i$ in a redefinition of ${\bf \hat s}$, and
dropping an infinite multiplicative constant (along with the Jacobian), we
have

\begin{eqnarray}
Z & = & \int
{\cal D}\hat s\, {\cal D} s\,\exp \left\{ \int dtd^dx\,{\bf \hat s}\cdot \left(
 -\eta \partial _t{\bf s}+K\nabla ^2{\bf s}+{\bf H}+{\bf h}-\frac{\partial
\tilde V}
{\partial {\bf s}}\right) \right\} \nonumber \\  & \equiv  & \int {\cal
D}\hat s\, {\cal D} s\,\exp \left({\cal S}\right) \text{.}
\end{eqnarray}
The {\it generating functional} $Z$ enables us to evaluate all correlation
and response functions. For example, the solution of Eq.~(\ref{motion}) is 
\begin{equation}
\label{sol}{\bf s}^{\text{sol}}\left( {\bf x},t\right) =\int {\cal D}\hat s\,
{\cal D}s\ {\bf s}\left( {\bf x},t\right) \exp \left( {\cal S}\right) \text{,}
\end{equation}
and its response to the magnetic field is given by 
\begin{equation}
\label{response}\frac{\delta s_1^{\text{sol}}\left( {\bf x},t\right) }{
\delta H\left(t^{\prime}\right)}=\int {\cal D}\hat s\,{\cal D}s
\int d^d{\bf x}^{\prime }\ \hat
s_1\left( {\bf x^{\prime }},t^{\prime}\right) s_1\left( {\bf x},t\right) 
\exp \left( {\cal S}\right) \text{.}
\end{equation}
Also, we can change the origin of time by a trivial reparametrization of
the magnetic field, as {\it e.g.} in
\begin{equation}
{\bf s}^{\text{sol}}\left( {\bf x},t+\epsilon;{\bf H}\left(t\right)\right) 
={\bf s}^{\text{sol}}\left( {\bf x},t;{\bf H}\left(t+\epsilon\right)\right)
\text{,}
\end{equation}
The latter expression takes the form ${\bf s}^{\text{sol}}\left( {\bf x},t;
{\bf H}\left(t\right)+{\bf \Omega}\epsilon\right)$ if $H$ is increased
linearly in time at a rate $\Omega$, whence
\begin{equation}
{\bf s}^{\text{sol}}\left( {\bf x},t+\epsilon\right) 
-{\bf s}^{\text{sol}}\left( {\bf x},t\right)
=\int {\cal D}\hat s {\cal D}s\, {\bf s}\left( {\bf x},t\right) 
\exp \left( {\cal S}\right) \left\{ \exp \left( \int dt^{\prime}
d^dx^{\prime}\,{ \hat s}_1 \Omega \epsilon \right) -1 \right\}
\text{,}
\end{equation}
from which it follows that the dynamic susceptibility is calculated as
\begin{equation}
\frac{\partial {\bf s}^{\text{sol}}\left( {\bf x},t\right) }{
\partial t}=\Omega \int dt^{\prime}d^dx^{\prime}\,
\int {\cal D}\hat s\,{\cal D}s \, \hat
s_1\left( {\bf x^{\prime }},t^{\prime }\right) {\bf s}
\left( {\bf x},t\right) \exp \left( 
{\cal S}\right) \text{.}
\end{equation}
Since we are interested in the average of the correlations and responses
over the random field, from now on we deal with the average of $Z$. This
enables us to forget the stochastic variable $h_\alpha $, trading it for a
new term in the ``averaged action''. Taking advantage of the Gaussian nature
of $h_\alpha $,

\begin{eqnarray}
\label{averaged} \overline{Z} & = & \int
{\cal D}\hat s\, {\cal D} s\,\exp \left\{ \int dtd^dx\,{\bf \hat s}\cdot \left(
 -\eta \partial _t{\bf s}+K\nabla ^2{\bf s}+{\bf H}-\frac{\partial \tilde
V}{\partial {\bf s}}
\right) \right\} \overline{\exp \int dtd^dx\,{\bf \hat s\cdot h}}
 \nonumber \\  & \equiv  & \int {\cal D}\hat s\, {\cal D}
s\,\exp\left(S\right) \text{,}
\end{eqnarray}
with (using Eq.~(\ref{distribution})) 
\begin{equation}
\label{action1}S=\int dtd^dx\,{\bf \hat s}\cdot \left( -\eta \partial _t{\bf s}
+K\nabla ^2{\bf s}+{\bf H}-\frac{\partial \tilde V}{\partial {\bf s}}
\right) +\int dtdt^{\prime }d^dx\,\frac R2{\bf \hat s}\left( {\bf x}
,t\right) \cdot {\bf \hat s}\left( {\bf x},t^{\prime }\right) \text{.}
\end{equation}

We have reformulated the theory, originally described by a dynamical
differential equation, in terms of an action $S\left[ {\bf s}\left( {\bf x}
,t\right) \right] $, which depends on the entire history of all spins. Thus 
$S$ is a functional of the path which the system follows; the probability
weight $\exp \left( S\right) $ picks the physical path and averages it over
disorder. We can then study the symmetries and renormalization of the
theory, as for equilibrium field theories.

The motion comprises two time scales: $\eta $ and $\left( dH/dt\right) ^{-1}$. 
The behavior of the system depends of course on the ratio of the two, and
not on their respective values. For the calculation of static exponents, we
let $\eta \rightarrow 0$. Consider for example the exponent $\nu $, with
which the correlation length diverges. A diverging correlation length gives
rise to an infinite susceptibility, detected by a non-vanishing response (of
the magnetization) to an infinitesimal increase of the magnetic field.
Clearly, such a behavior is obtained in our problem only if the magnetic
field increases infinitely slowly [{\it i.e.} $\left( dH/dt\right)
\rightarrow 0$], or equivalently, if $\eta \rightarrow 0$. As $\eta $ is a
measure of how much the system lags behind its local energy minimum, for 
$\eta \rightarrow 0$ the system always stays at the time dependent minimum
(as seen by setting $\eta =0$ in the equation of motion). In other words,
the spin avalanches resulting from a small change in $H$ spread
instantaneously.

In the $\eta \rightarrow 0$ limit, time evolution of the spin field is
simply motion with the minimum in the energy landscape. There is
nevertheless a subtlety involved as to the choice of the minimum. For
example, because of the double-well shape of the potential, a scalar spin
has the choice, during some time interval of its history, between two
positions, both of which locally minimize the Hamiltonian. If we solve the
equation of motion, there is no ambiguity, since we specify one of the
energy minima as the initial condition, and then follow its time evolution.
In particular, the initial condition corresponding to the case of a magnetic
field which increases from $-\infty $ to $+\infty $ is one in which all the
spins occupy the left minimum of their double wells \cite{initialcond}. In
the path integral formalism, however, no initial condition is specified. The
weight $e^S$ picks all possible solutions, corresponding to different
initial conditions. Let us illustrate this point with, as in
Ref.~\cite{parisi}, the zero-dimensional non-random version of our model,
defined by the equation of motion 
\begin{equation}
\eta \partial _ts=-\frac \partial {\partial s}\left( -Hs+as^2+bs^4\right) 
\text{,}
\end{equation}
with $\eta \rightarrow 0$. We can imagine $s$ as a bead sitting at the
minimum of the quartic potential $Q\left( s\right) =-Hs+as^2+bs^4$, and
moving with it. With $a<0$ and $b>0$, $Q\left( s\right) $ has a single
minimum if $\left| H\right| $ is larger than some value $\Delta $, and two
minima if $\left| H\right| $ is smaller than $\Delta $. For $H$ very
negative, the bead sits at the single minimum of the curve, which becomes
the left minimum when $H$ is between $-\Delta $ and $\Delta $. At $H=\Delta $, 
the left minimum disappears and the bead moves to the right minimum. Thus,
in particular, $s\left( -H\right) \neq -s\left( H\right) $ and $s\left(
H=0\right) \neq 0$: the motion of $s$ {\it is hysteretic} (Fig.~2).

\begin{figure} 
\epsfxsize=10truecm 
\vspace*{.6truecm} 
\centerline{
\epsfbox{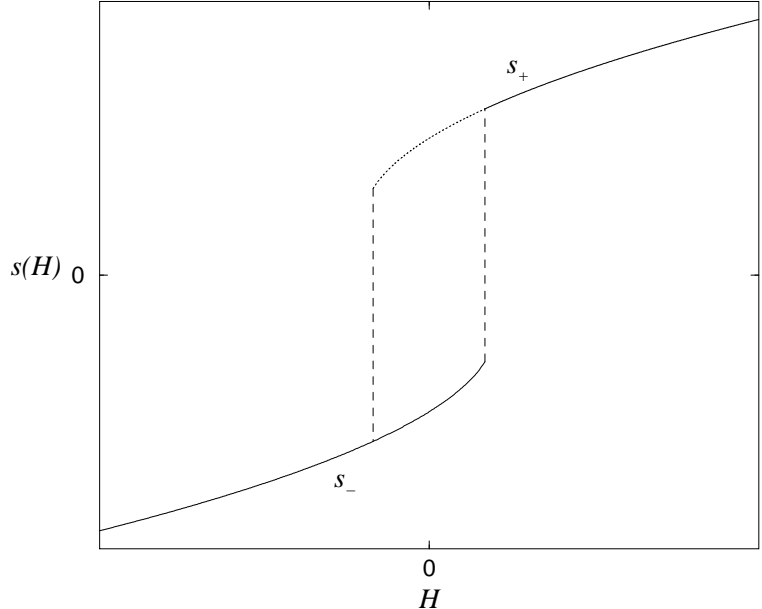} 
} 
\vspace*{.3truecm}

\caption{Metastable solutions of the single-spin system. The physical
solution for an increasing magnetic field (solid line) follows the lower
curve ($s_-$) as long as it is present, then jumps along the dashed line
to the upper curve ($s_+$).}
\vspace*{.6truecm}
\end{figure}
The action for this model is 
\begin{equation}
S=\int dt\,\hat s\left( -\eta \partial _ts+H-as-bs^3\right) .
\end{equation}
Let {\it $\left\langle s\right\rangle $} be the average of $s$ with respect
to the weight $e^S$. Calculating {\it \ $\left\langle s\right\rangle $}
perturbatively, it is easily seen (by counting the possible occurrences of $s$
and $\hat s$ in a diagram) that each diagram comes with an odd number of $H$'s. 
Therefore, {\it $\left\langle s\left( -H\right) \right\rangle
=-\left\langle s\left( H\right) \right\rangle $}, and in particular, 
{\it $\left\langle s\left( H=0\right) \right\rangle =0$}, in apparent
contradiction with the above solution. But as mentioned earlier, the
partition function is merely the integral of a delta function which imposes
the equation of motion. In the present case, it imposes 
\begin{equation}
\frac \partial {\partial s}\left( -Hs+as^2+bs^4\right) =0.
\end{equation}
Thus, in calculating {\it $\left\langle s\right\rangle $}, we pick all
minima of the quartic form. In other words, {\it $\left\langle
s\right\rangle $} is the sum of two terms, {\it $\left\langle s\right\rangle
=\left\langle s_{-}\right\rangle +\left\langle s_{+}\right\rangle $},
corresponding to the left and right minima. The physical, ``bead''
solution is equal to {\it $\left\langle s_{-}\right\rangle $} up to 
$H=\Delta $ and then {\it $\left\langle s_{+}\right\rangle $} for $H$ larger
than $\Delta $. Similarly, in the case of our original problem, the quantity 
${\bf s}^{\text{sol}}\left( {\bf x},t\right) $ of Eqs.~(\ref{sol}) and (\ref{response}) 
does not coincide with the physical solution we are looking for.
In addition to the latter, ${\bf s}^{\text{sol}}\left( {\bf x},t\right) $
contains other unwanted solutions corresponding to additional energy minima.
We shall come back to this difficulty and circumvent it in the next section,
both for the scalar and the vector models.

\section{Perturbative renormalization}

\subsection{Coordinates and Fields Rescalings, the Free Propagator}

Our renormalization group transformation consists of the usual three steps.
First, we coarse-grain the system, {\it i.e.}, integrate out the modes with
wave number between $\Lambda /b$ and $\Lambda $ ($b>1$). Second, we rescale
coordinates ({\it i.e.} change our length and time units), by setting 
\begin{equation}
{\bf x}\rightarrow b{\bf x}\text{,\quad }t\rightarrow b^zt\text{,} 
\end{equation}
or, equivalently 
\begin{equation}
{\bf q}\rightarrow b^{-1}{\bf q}\text{,\quad }\omega \rightarrow
b^{-z}\omega \text{.} 
\end{equation}
With this change of units, the coarse-grained fields vary on the same length
scales as the original ones, and the lattice cutoff is preserved. Third, we
rescale the fields according to 
\begin{equation}
{\bf s}\left( b{\bf x},b^zt\right) \rightarrow \zeta {\bf s}\left( {\bf x}
,t\right) \text{,\quad }{\bf \hat s}\left( b{\bf x},b^zt\right) \rightarrow
\hat \zeta {\bf \hat s}\left( {\bf x},t\right) \text{,} 
\end{equation}
or, equivalently 
\begin{equation}
{\bf s}\left( b^{-1}{\bf q},b^{-z}\omega \right) \rightarrow b^{d+z}\zeta 
{\bf s}\left( {\bf q},\omega \right) \text{,\quad }{\bf \hat s}\left( b^{-1}
{\bf q},b^{-z}\omega \right) \rightarrow b^{d+z}\hat \zeta {\bf \hat s}
\left( {\bf q},\omega \right) \text{.} 
\end{equation}
With the choice

\begin{eqnarray}
z & = & 2
\text{,} \\ \zeta & = & b^{2-\frac d2}
\text{,} \\ \hat \zeta & = & b^{-2-\frac d2}\text{,}
\end{eqnarray}
the time derivative and Laplacian terms of the action, as well as the $\hat
s\hat s$ terms, become scale invariant. The recursion relations will be
calculated to the lowest non-trivial order in the interaction. To this
order, no corrections of the parameters $\eta $, $K$, or $R$ of the action
in Eq.~(\ref{action1}) occur in the coarse-graining transformation, {\it i.e.} 
$\eta $, $K$, and $R$ are invariant under our perturbative renormalization.

As in the momentum space RG treatment of the $\phi ^4$ theory, we consider
the quadratic part of the action as a Gaussian (free) theory, and the rest
as a perturbation (interaction). Following Refs.~\cite{dahmen,prb,thesis}, for
calculational convenience we treat the disorder-induced $\hat s\hat s$ term
as an interaction, instead of including it in the Gaussian part. The free
theory thus consists only of the $\hat ss$ part of the action, and the
corresponding bare propagators are 
\begin{equation} 
\left\{ 
\begin{array}{lll}
\left\langle s_\alpha \left( {\bf q},\omega \right) s_\beta \left( {\bf 
q^{\prime }},\omega ^{\prime }\right) \right\rangle _0 & = & 0 \\ 
\left\langle \hat s_\alpha \left( {\bf q},\omega \right) \hat s_\beta \left( 
{\bf q^{\prime }},\omega ^{\prime }\right) \right\rangle _0 & = & 0 \\ 
\left\langle \hat s_\alpha \left( {\bf q},\omega \right) s_\beta \left( {\bf 
q^{\prime }},\omega ^{\prime }\right) \right\rangle _0 & = & \delta _{\alpha
\beta }2\pi \delta \left( \omega +\omega ^{\prime }\right) \left( 2\pi
\right) ^d\delta ^d\left( {\bf q}+{\bf q}^{\prime }\right) \frac 1{-\eta
i\omega +Kq^2-r_\alpha } 
\end{array}
\right. \text{,} 
\end{equation}
where $\left\langle \cdot \right\rangle _0$ denotes an average with respect
to the Gaussian weight, and the indices run from 1 to $n$. The parameter 
$r_\alpha $ is the $q$-independent coefficient of the quadratic $\hat
s_\alpha s_\alpha $ term in the action. Fourier transforming back in time,
we have 
\begin{equation}
\left\langle \hat s_\alpha \left( {\bf q},t\right) s_\beta \left( {\bf 
q^{\prime }},t^{\prime }\right) \right\rangle _0=\left\{ 
\begin{array}{l}
0 
\text{\quad if\quad }t^{\prime }\leq t \\ \delta _{\alpha \beta }\left( 2\pi
\right) ^d\delta ^d\left( {\bf q}+{\bf q}^{\prime }\right) \exp \left[ - 
\frac{Kq^2-r_\alpha }\eta \left( t^{\prime }-t\right) \right] /\eta \text{
\quad if\quad }t^{\prime }>t 
\end{array}
\right. \text{.} 
\end{equation}
(This is calculated for $r_\alpha <0$. As we shall see, $r_\alpha $ is
indeed negative at criticality.) In the $\eta \rightarrow 0$ limit, the
propagator becomes 
\begin{equation}
\left\langle \hat s_\alpha \left( {\bf q},t\right) s_\beta \left( {\bf 
q^{\prime }},t^{\prime }\right) \right\rangle _0=\delta _{\alpha \beta
}\left( 2\pi \right) ^d\delta ^d\left( {\bf q}+{\bf q}^{\prime }\right)
\frac 1{Kq^2-r_\alpha }\delta \left( t^{\prime }-t^{+}\right) \text{.} 
\end{equation}
That is, the contraction of $\hat s_\alpha \left( t\right) $ and $s_\alpha
\left( t^{\prime }\right) $ is non-vanishing only if the two times are equal
(actually, only if $t^{\prime }$ is infinitesimally higher than $t$). With
this propagator, it is easily seen diagrammatically that, although the
disorder-induced $\hat s\hat s$ terms couple different times, a renormalized
vertex at time $t$ is a function only of the other vertices {\it at the same
time }$t$. Thus, slices of the action at different times renormalize
independently from each other, and flow to their respective fixed points.
This justifies the procedure of Refs.~\cite{dahmen,prb,thesis} of setting $H$
constant for the calculation of static exponents. (In what follows, we shall
need to correct the action of Eq.~(\ref{action1}) to take care of the
problem of multiple solutions. We shall, for example, expand $S$ about a
uniform but time-dependent value of the field, say some function ${\bf 
\sigma }\left( t\right) $. Hence the vertices (coefficients in $S$), and in
particular the ``masses'' $r_\alpha $, become functions of the parameter 
${\bf \sigma }\left( t\right) $. Note that the above analysis of the $\eta
\rightarrow 0$ limit is then legitimate only if ${\bf \sigma }\left(
t\right) $ is continuous, so that ${\bf \sigma }\left( t^{+}\right) ={\bf 
\sigma }\left( t\right) $. Below, we shall define our ${\bf \sigma }$'s in
terms of the magnetization or analogous quantities. Thus, our analysis holds
if we approach criticality from the high disorder side.)

\subsection{The Scalar Model Revisited}

The zero-dimensional model discussed in Sec. III is nothing but the
single-spin equivalent to the scalar field. The apparent contradiction
mentioned therein is also present in the full model. Let us call $s{\bf _{-}}
\left( {\bf x},t\right) $ and $s{\bf _{+}}\left( {\bf x},t\right) $ the
solutions of Eq.~(\ref{motion}), for a magnetic field $H$ increasing from 
$-\infty $ to $+\infty $, or decreasing from $+\infty $ to $-\infty $,
respectively. The magnetization measured experimentally is the average over
space, or equivalently over the random field, of the above solutions, 
\begin{equation}
m_{\pm }\left( H\left( t\right) \right) =\overline{s{\bf _{\pm }}\left( {\bf 
x},t\right) }\text{.} 
\end{equation}
Generically, as is inferred from the single-spin case and observed
experimentally, the magnetization displays hysteresis. In particular, 
$m_{\pm }\left( -H\right) \neq -m_{\pm }\left( H\right) $ and $m_{\pm }\left(
H=0\right) \neq 0$. On the other hand, the action of Eq.~(\ref{action1}) is
invariant under the transformation $\left( \hat s,s,H\right) \rightarrow
\left( -\hat s,-s,-H\right) $. This implies that the average of $s^{\text{sol}}$ 
[Eq.~(\ref{sol})] satisfies $\overline{s^{\text{sol}}\left( -H\right) }
=- \overline{s^{\text{sol}}\left( H\right) }$, and $\overline{s^{\text{sol}}
\left( H=0\right) }=0$. As explained above, $s^{\text{sol}}$ contains
unphysical contributions (corresponding to the many minima in the energy
landscape) in addition to the physical solution ($s{\bf _{-}}$ in the case
of an increasing $H$), {\it i.e.} 
\begin{equation}
s^{\text{sol}}\left( {\bf x},t\right) =s{\bf _{-}}\left( {\bf x},t\right) +s
{\bf ^{\prime }}\left( {\bf x},t\right) +s{\bf ^{\prime \prime }}\left( {\bf 
x},t\right) +\cdots \text{.} 
\end{equation}
Hence the action 
\begin{equation}
\label{action2}S=\int dtd^dx\,\hat s\left( -\eta \partial _ts+K\nabla
^2s+H+as+bs^3+\cdots \right) +\int dtdt^{\prime }d^dx\,\frac R2\hat s\left( 
{\bf x},t\right) \hat s\left( {\bf x},t^{\prime }\right) \text{,} 
\end{equation}
does not describe the magnetization, unless it is corrected in such a way as
to remove the unphysical solutions. These are inopportunly introduced 
through Eq.~(\ref{delta}), which should in fact be written as

\begin{eqnarray}
& & \int 
{\cal D}s\, \delta \left( -\eta \partial _ts+K\nabla ^2 s+ H+
h-\frac{\partial \tilde V}{\partial s}\right) \times \left( 
\text{Jacobian}\right) \nonumber 
\\ & & =\int {\cal D}s\, \left\{ \delta \left( s-s
_{-}\left( {\bf x},t\right) \right) +\delta \left( s-s^{\prime
}\left( {\bf x},t\right) \right) +\delta \left( s- s^{\prime
\prime }\left( {\bf x},t\right) \right) +\cdots \right\}.
\end{eqnarray}
Therefore, if we substract the quantity $\overline{\delta \left( s-
s^{\prime }\right) +\delta \left( s-s^{\prime \prime
}\right) +\cdots }$ from the functional $e^S$, we obtain a well defined
theory, which incorporates only the physical solution and yields all the
correct correlation and response functions. Although this method is
impossible to implement (since it explicitely involves the unphysical
solutions), a restricted, weaker version is applicable, and fully serves our
purposes. The investigation of the critical behavior and the calculation of
the corresponding exponents relies primarily on correlation and response
functions, namely the field density $s$ and its susceptibility $\partial 
s / \partial t$, {\it which are linear in} $s$. For such objects, 
$\delta \left( s-s_{-}\right) +\delta \left( s- s
^{\prime }\right) +\delta \left( s- s^{\prime \prime }\right)
+\cdots $ can be replaced with $\delta \left(s- s
_{-}- s^{\prime }- s^{\prime \prime }-\cdots \right) $.
Furthermore, only averaged quantities are of interest to the study of the
critical point, and linearity allows to perform
the average inside the argument of the delta
function, leading to
\begin{equation}
\delta \left( s- m_{-}\left( t\right) -\sigma \left(
t\right) \right) \text{,} 
\end{equation}
where
\begin{equation}
\sigma \left( t\right) =\overline{ s^{\prime }\left( {\bf x}
,t\right) + s^{\prime \prime }\left( {\bf x},t\right) +\cdots }\text{.} 
\end{equation}
A comparison with Eqs.~(\ref{delta}) and (\ref{averaged}) 
implies that a weight $e^S$ which
properly describes the averaged theory is obtained by adding $\sigma \left(
t\right) $ to the argument of the action. Indeed, for 
\begin{equation}
\left\langle s\right\rangle \equiv \int {\cal D}\hat s\,{\cal D}s\,s\left( 
{\bf x},t\right) e^{S\left[ \hat s,s\right] }\text{,} 
\end{equation}
we have $\left\langle s\right\rangle =m_{-}\left( t\right) +\sigma \left(
t\right) $; therefore shifting the field by $\sigma $$\left( t\right) $
yields 
\begin{equation}
\int {\cal D}\hat s\,{\cal D}s\,s\left( {\bf x},t\right) e^{S\left[ \hat
s,s+\sigma \right] }=\int {\cal D}\hat s\,{\cal D}s\,\left[ s{\bf \left(
x,t\right) }-\sigma \left( t\right) \right] e^{S\left[ \hat s,s\right]
}=\left\langle s\right\rangle -\sigma \left( t\right) =m_{-}\left( t\right) 
\text{.} 
\end{equation}
The average spin dynamics is thus properly described by the corrected action

\begin{eqnarray}
\label{scalaction}S & = & S\left[ \hat s,s+\sigma \right] \nonumber \\  
& = & \int dtd^dx\,\hat s\left( -\eta \partial _ts+K\nabla
^2s+H+A_0+A_1s+A_2s^2+A_3s^3+\cdots \right) +\int dtdt^{\prime }d^dx\,\frac
R2\hat s\left( {\bf x},t\right) \hat s\left( {\bf x},t^{\prime }\right) 
\text{,}
\end{eqnarray}
where 
\begin{equation}
\left\{ 
\begin{array}{lll}
A_0 & = & -\eta \partial _t\sigma +a\sigma +b\sigma ^3+\cdots \\ 
A_1 & = & a+3b\sigma ^2+\cdots \\ 
A_2 & = & 3b\sigma +\cdots \\ 
A_3 & = & b+\cdots \\ 
\vdots &  &  
\end{array}
\right. \text{.} 
\end{equation}

Although we cannot obtain the precise form of $\sigma \left( t\right) $
without solving for the many minima, its qualitative shape is easily found.
Consider a very large (positive or negative) magnetic field. The unphysical
minima are due only to the spins in a very large random field $h$, such that
the potential they feel still has two minima. When we take the average, the
minima $s{\bf ^{\prime }}+s{\bf ^{\prime \prime }}+\cdots $ contribute to 
$\sigma $ only with a very small weight [$\propto \exp \left( -h^2/2R\right) $], implying 
\begin{equation}
\sigma \rightarrow 0^{\pm }\text{, as }H\rightarrow \pm \infty \text{.} 
\end{equation}
Furthermore, it follows from the definition of $\sigma $ that $\sigma
=-m_{-} $ at $H=0$. Thus, $\sigma \left( H\right) $ has a bell shape,
shifted to the right since the sum of $\sigma $ and $m_{-}$ is an odd
function of $H$ (Fig.~3). The parameter $\sigma \left( t\right) $, which
corrects the coefficients in $S$, breaks the up-down symmetry present in the
action of Eq.~(\ref{action2}). This is physically required, since in an
hysteretic system, the (non-equilibrium) magnetization breaks that symmetry,
in particular at $H=0$.

\begin{figure} 
\epsfxsize=12truecm 
\vspace*{.6truecm} 
\centerline{
\epsfbox{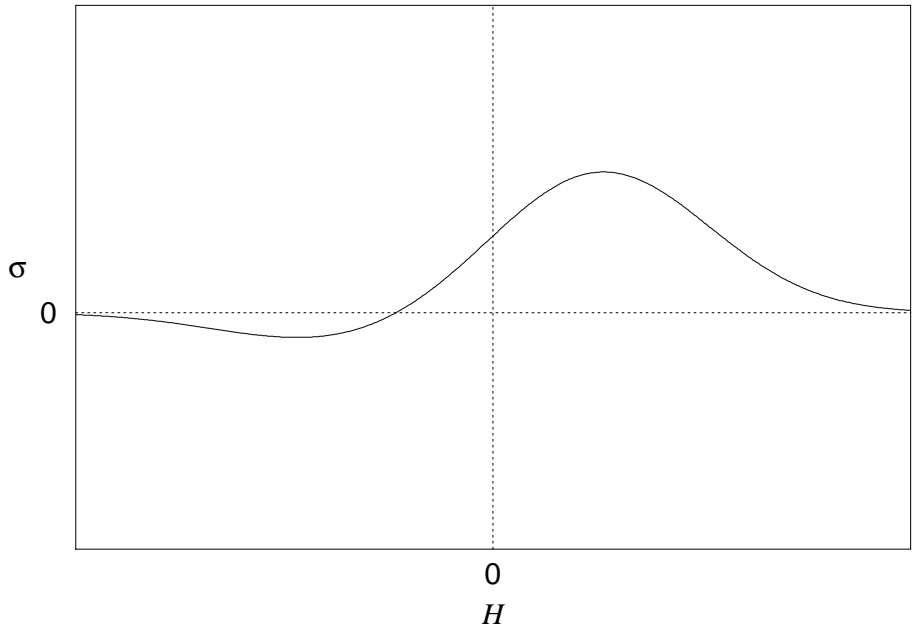} 
} 
\vspace*{.3truecm}

\caption{Schematic shape of the parameter $\sigma$ as a function of the
magnetic field.}
\vspace*{.6truecm}
\end{figure}

To discuss the relevant terms in the action, we apply a renormalization
transformation in $6-\epsilon $ dimensions. All terms of the action higher
than cubic in $s$ are irrelevant, and after a large enough number of
coarse-graining steps, we are left with the renormalized action 
\begin{equation}
\label{renoraction}S=\int dtd^dx\,\hat s\left( -\eta \partial _ts+K\nabla
^2s+\tilde A_0+\tilde A_1s+\tilde A_2s^2+\tilde A_3s^3\right) +\int
dtdt^{\prime }d^dx\,\frac R2\hat s\left( {\bf x},t\right) \hat s\left( {\bf x%
},t^{\prime }\right) \text{,}
\end{equation}
where $\tilde A_{i\geq 1}$'s are functions of $H$ through $\sigma $, and $%
\tilde A_3$ is negative to prevent the spins from diverging in magnitude.
Furthermore, it is easily shown that $\tilde A_i$ is an even (odd) function
of $\sigma $ for $i$ odd (even). In particular, since $\sigma \rightarrow 0$
as $H\rightarrow \pm \infty $, we also have that $\tilde A_2\rightarrow 0$
as $H\rightarrow \pm \infty $. Now, in order to find the critical point, let
us expand the field about some value $\mu \left( t\right) $, so as to cancel 
$\tilde A_2$. With the choice $\mu =-\tilde A_2/3\tilde A_3$, the action in
terms of $s^{\prime }\equiv s-\mu $ becomes 
\begin{equation}
S=\int dtd^dx\,\hat s\left( -\eta \partial _ts^{\prime }+K\nabla ^2s^{\prime
}+\tilde A_0^{\prime }+\tilde A_1^{\prime }s^{\prime }+\tilde A_3^{\prime
}s^{\prime 3}\right) +\int dtdt^{\prime }d^dx\,\frac R2\hat s\left( {\bf x}%
,t\right) \hat s\left( {\bf x},t^{\prime }\right) \text{,}
\end{equation}
where the coefficients have been corrected by $\mu $. Since $\left\langle
s\right\rangle \rightarrow \pm \infty $ as $H\rightarrow \pm \infty $, $%
\left\langle s\right\rangle $ crosses $\mu $ at a given $H_0$, {\it i.e.} $%
\left\langle s\left( H_0\right) \right\rangle =\mu \left( H_0\right) $.
Hence $\left\langle s^{\prime }\left( H_0\right) \right\rangle =0$, which
implies in general that $\tilde A_0^{\prime }\left( H_0\right) =0$. The
action thus reduces to that studied in Refs.~\cite{dahmen,prb,thesis}, and is
critical for a specific amount of disorder $R$.

\subsection{The vector model}
\subsubsection{The appropriate action}

The first question in the case of vector spins is whether the action of
Eq.~(\ref{action1}) correctly describes the system. The single-spin problem
is identical to that of a bead moving in a Mexican hat tilted by ${\bf H}+
{\bf h}$, the sum of the external ${\bf H}\left( t\right) $ and the quenched
random field ${\bf h}$. The bead simply turns around the bump of the hat, 
{\it i.e.} the spin always points in the direction of ${\bf H}+{\bf h}$, as
in the equilibrium problem. The motion is governed by a single minimum, and
there is no hysteresis. Similarly, mean field theory reduces the time
evolution of the spin field to that of a single degree of freedom, and
yields no hysteresis for any value of the disorder. These observations may
suggest that Eq.~(\ref{action1}) properly describes the problem, and that
criticality occurs at ${\bf H}=0$, simultaneously for the components of the
field parallel and perpendicular to ${\bf H}$. However, although a single
spin has only one minimum in its energy landscape, a configuration of
several spins may have many. Consider a system composed of two spins of unit
length, in an increasing magnetic field ${\bf H}$, and with random fields 
$+{\bf \Delta }$ and $-{\bf \Delta }$ perpendicular to ${\bf H}$, as
illustrated on Fig.~4.

\begin{figure} 
\epsfxsize=9truecm 
\vspace*{.6truecm} 
\centerline{
\epsfbox{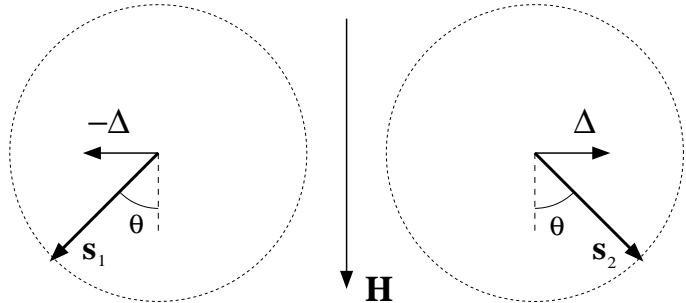} 
} 
\vspace*{.3truecm}

\caption{Illustration of the two-spin toy model described in the text.}
\vspace*{.6truecm}
\end{figure}
The Hamiltonian for this system is 
\begin{equation}
{\cal H}=-J{\bf s}_1\cdot {\bf s}_2-{\bf \Delta \cdot }\left( {\bf s}_1-{\bf 
s}_2\right) -{\bf H}\cdot \left( {\bf s}_1+{\bf s}_2\right) \text{.} 
\end{equation}
If $\theta _1$ and $\theta _2$ are the angles of the spins with respect to 
${\bf H}$, the minimum energy path that the spins follow imposes 
$\theta_1=-\theta _2\equiv \theta $, and in terms of $\theta $, the energy is 
\begin{equation}
{\cal H}=2\left\{ J\left( \sin \theta \right) ^2-\Delta \sin \theta +H\cos
\theta \right\} -J\text{.} 
\end{equation}
For $\Delta <2J$ and $H=0$, the energy as a function of $\theta $ is a
symmetric double well centered on $\theta =\pi /2$ (Fig.~5), similar to the
scalar single-spin energy landscape, with two minima at $\sin \theta =\Delta
/2J$. A non-vanishing $H$ tilts the double well, and ultimately suppresses
one of the two minima. The ferromagnetic interaction causes the two spins to
pull each other back before jumping ahead, thereby investing the $n$-component
field's motion with a hysteretic, uniaxial-like character. That is, the two-spin
toy system as a whole goes over an energy barrier, reminiscent of the motion
of a scalar spin and in contrast with that of a single multicomponent spin which
turns around the energy barrier. Interestingly, the Ising (hysteretic)
behavior of the {\it longitudinal} component results from the 
interaction, and the presence of a {\it transverse} random field.

\begin{figure} 
\epsfxsize=10truecm 
\vspace*{.6truecm} 
\centerline{
\epsfbox{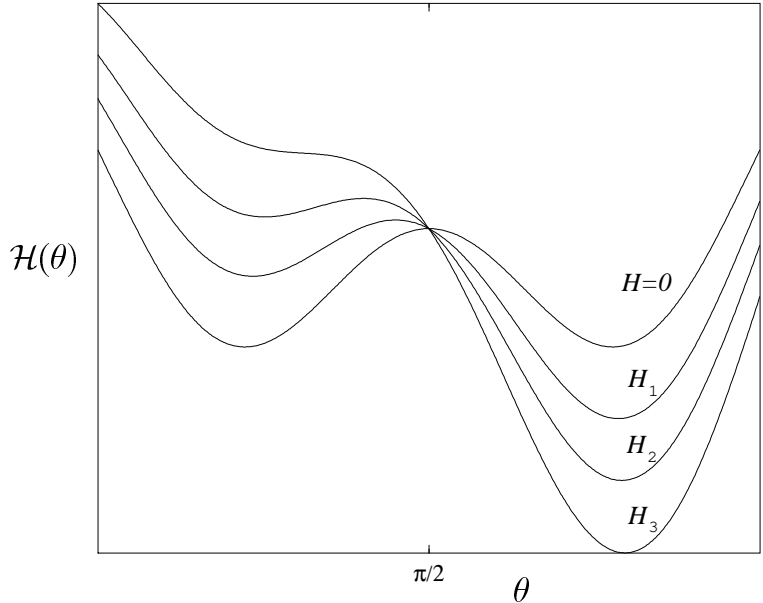} 
} 
\vspace*{.3truecm}

\caption{Energy profile of the two-spin toy model, for different values
of the magnetic field $H$. At $H=0$, the double well is symmetric.
Larger values of $H$ tilt the curve, eventually suppressing the left
minimum.}
\vspace*{.6truecm}
\end{figure}

In the zero dimensional (single spin) or infinite dimensional (mean field)
case, the system follows a single minimum and there is no hysteresis. When
fluctuations are included, however, the above toy system shows that many
minima appear, allowing for a hysteretic behavior. Thus, as in the scalar
case, the action of Eq.~(\ref{action1}) has to be corrected in order to
remove the unphysical minima. Following the procedure of Sec.~IV.B, we
replace ${\bf s}\left( {\bf x},t\right) $ by ${\bf s}\left( {\bf x},t\right)
+{\bf \sigma }\left( t\right) $, to obtain, if ${\bf \sigma }$ is parallel
to ${\bf H}$, the corrected action
\begin{eqnarray}
S & = & S\left[ 
{\bf \hat s},{\bf s}+{\bf \sigma }\right] \nonumber \\  & = & \int dtd^dx\,\hat
s_{\Vert }\left\{ -\eta \partial _t\left( s_{\Vert }+\sigma \right) +K\nabla
^2s_{\Vert }+H+c_1\left( s_{\Vert }+\sigma \right) +c_2\left[ \left(
s_{\Vert }+\sigma \right) ^2+
{\bf s}_{\bot }^2\right] \left( s_{\Vert }+\sigma \right) +\cdots \right\} 
\nonumber \\  &  & +\int dtdt^{\prime }d^dx\,\frac R2\hat s_{\Vert }\left( 
{\bf x},t\right) \hat s_{\Vert }\left( {\bf x},t^{\prime }\right) \nonumber
 \\  &  & +\int dtd^dx\,
\hat {\bf s}_{\bot }\cdot \left\{ -\eta \partial _t{\bf s}_{\bot }+K\nabla ^2
{\bf s}_{\bot }+c_1{\bf s}_{\bot }+c_2\left[ \left( s_{\Vert }+\sigma
\right) ^2+{\bf s}_{\bot }^2\right] {\bf s}_{\bot }+\cdots \right\} 
\nonumber \\  &  
& +\int dtdt^{\prime }d^dx\,\hat {\bf s}_{\bot }\left( {\bf x},t\right)
\cdot \hat {\bf s}_{\bot }\left( {\bf x},t^{\prime }\right), \label{compact}
\end{eqnarray}
where the subscripts $\Vert $ and $\bot $ refer to the components parallel
and perpendicular to the magnetic field. Expanding the polynomials, and
absorbing the parameter $\sigma $ in thereby corrected coefficients, leads
to the form
\begin{eqnarray}
S & = & \int dtd^dx\, \hat s_{\Vert }\left( -\eta \partial _ts_{\Vert
}+K\nabla ^2s_{\Vert }+A_0+A_1s_{\Vert }+A_2s_{\Vert }^2+\bar A_2
{\bf s}_{\bot }^2+A_3s_{\Vert }^3+\bar A_3{\bf s}_{\bot }^2s_{\Vert }
+\cdots \right) \nonumber  \\  
&  & +\int dtdt^{\prime }d^dx\,\frac R2\hat s_{\Vert }\left( 
{\bf x},t\right) \hat s_{\Vert }\left( {\bf x},t^{\prime }\right) \nonumber
 \\  &  & 
+\int dtd^dx\, 
\hat {\bf s}_{\bot }\cdot \left( -\eta \partial _t{\bf s}_{\bot }+K\nabla
^2%
{\bf s}_{\bot }+B_1{\bf s}_{\bot }+B_2s_{\Vert }{\bf s}_{\bot }+B_3{\bf s}_{\bot
}^2{\bf s}_{\bot }+\bar B_3s_{\Vert }^2{\bf s}_{\bot }+\cdots \right)
\nonumber \\  &  & +\int dtdt^{\prime }d^dx\,\frac R2 \hat {\bf s}_{\bot }
\left( {\bf x},t\right) \cdot \hat {\bf s}_{\bot }\left( {\bf x},t^{\prime
}\right) \text{.} \label{expanded}
\end{eqnarray}
Clearly, only even powers of $s_{\bot }$ are present in the first line,
while only odd powers occur in the third. A term of order $p$ in ${\bf s}$,
trivially ({\it i.e.} ignoring the coarse graining step which couples
different order terms) scales with $b^{2+d}\hat \zeta \zeta ^p=b^{2p-\left(
p-1\right) d/2}$. In dimensions $d\leq 4$, all terms are relevant, and a
non-trivial fixed point at long length scales cannot in general be reached
by tuning merely two quantities, namely the external magnetic field and the
amount of randomness. Similarly for $d=5$, in which terms up to $p=5$ are
relevant. In $6-\epsilon $ dimensions, however, all terms with $p>3$ are
irrelevant under the RG, leading to an effective action of the form of 
Eq.~(\ref{expanded}), with any terms not displayed set to zero. Including the
corrections due to coarse graining, to lowest non-trivial order (in the
interaction and in $\epsilon =6-d$), the recursion relations for the
remaining 10 vertices read 
\begin{equation}
\label{recursion}\left\{ 
\begin{array}{lll}
A_0^{\prime } & = & b^{3-\epsilon /2}\left[ A_0+\left( I+2A_1\right)
A_2+\left( n-1\right) \left( I+2B_1\right) \bar A_2\right]  \\ 
A_1^{\prime } & = & b^2\left[ A_1+3\left( I+2A_1\right) A_3+\left(
n-1\right) \left( I+2B_1\right) \bar A_3\right]  \\ 
B_1^{\prime } & = & b^2\left[ B_1+\left( n+1\right) \left( I+2A_1\right)
B_3+\left( I+2B_1\right) \bar B_3\right]  \\ 
A_2^{\prime } & = & b^{1+\epsilon /2}\left[ A_2+18A_2A_3+2\left( n-1\right)
\bar A_2\bar B_3+2\left( n-1\right) B_2\bar A_3\right]  \\ 
\bar A_2^{\prime } & = & b^{1+\epsilon /2}\left[ \bar A_2+2A_2\bar
A_3+2\left( n+1\right) \bar A_2B_3+2B_2\bar A_3+4\bar A_2\bar A_3\right]  \\ 
B_2^{\prime } & = & b^{1+\epsilon /2}\left[ B_2+4B_2\bar B_3+2\left(
n+1\right) B_2B_3+2B_2\bar A_3+4A_2\bar B_3+4\bar A_2\bar B_3\right]  \\ 
A_3^{\prime } & = & b^\epsilon \left[ A_3+18A_3^2+2\left( n-1\right) \bar
A_3\bar B_3\right]  \\ 
\bar A_3^{\prime } & = & b^\epsilon \left[ \bar A_3+4\bar A_3^2+6A_3\bar
A_3+2\left( n+1\right) \bar A_3B_3+4\bar A_3\bar B_3\right]  \\ 
B_3^{\prime } & = & b^\epsilon \left[ B_3+2\left( n+7\right) B_3^2+2\bar
A_3\bar B_3\right]  \\ 
\bar B_3^{\prime } & = & b^\epsilon \left[ \bar B_3+4\bar B_3^2+2\left(
n+1\right) B_3\bar B_3+6A_3\bar B_3+4\bar A_3\bar B_3\right] 
\end{array}
\right. ,
\end{equation}
where $I=\Lambda ^2\left( b^2-1\right) /2b^2\ln b$, and a factor of $R\ln
b/\left( 4\pi \right) ^3$ is absorbed in a redefinition of $A_2$, $\bar A_2$, 
$B_2$, $A_3$, $\bar A_3$, $B_3$, and $\bar B_3$. (For the derivation of
the corresponding relations for the scalar model, the reader is referred to
Refs.~\cite{prb,thesis}; the extension to the effective action of 
Eq.~(\ref{expanded}) is straightforward.) 

\subsubsection{Ising Criticality}
Clearly, a non-trivial fixed point for
all 10 vertices cannot be reached in general if only two quantities are to
be tuned. However, if $R$ is appropriately tuned, $A_1$ flows to its fixed
(finite) value, while $B_1$ grows indefinitely. Then, under the RG, the
interactions in the perpendicular components become less and less important
with respect to the quadratic term. After sufficient rescalings, the theory
becomes Gaussian in the perpendicular fields which can be integrated out,
resulting in an effective action for $s_{\Vert }$ identical to the scalar
action of Eq.~(\ref{scalaction}). Thus, the critical point of our ${\cal O}
\left( n\right) $ model is generically described by the same action as in
Refs.~\cite{dahmen,prb,thesis}, yielding identical recursion relations and
exponents. This can be physically understood in the following way: if a
configuration of several spins gives rise to many minima, a coarse-grained
vector-spin system roughly looks like an Ising system, leading to
scalar-like critical behavior. The latter occurs for a non-vanishing value
of the magnetic field or the magnetization, at which the full rotational
symmetry of the ${\cal O}\left( n\right) $ model is broken.

\subsubsection{Transverse criticality}

In the case considered above, the longitudinal field is massless and the
transverse components massive. Alternatively, one should be able to reach
another fixed point which incorporates the reverse situation. In 
Eq.~(\ref{expanded}), let us expand the longitudinal field about some parameter 
$\lambda \left( t\right) $, so chosen as to cancel the linear term $A_0$, 
{\it i.e.} we write 
\begin{equation}
s_{\Vert }=s_{\Vert }^{\prime }+\lambda \text{,} 
\end{equation}
with $\lambda $ satisfying 
\begin{equation}
\label{lambdaeq}A_0+A_1\lambda +A_2\lambda ^2+A_3\lambda ^3=0\text{.} 
\end{equation}
The theory is then Gaussian in $s_{\Vert }^{\prime }$. Integrating out the
longitudinal field, the action {\it for the transverse field} reduces to
Eq.~(\ref{renoraction}) with $\left({\bf \hat s}_{\bot },{\bf s}_{\bot }\right)$
instead of $\left(\hat s, s\right)$, and $\tilde A_0=\tilde A_2=0$, {\it i.e.} 
\begin{equation}
\label{transaction}S=\int dtd^dx\,{\bf \hat s}_{\bot }\cdot \left[ \left(
-\eta \partial _t+K\nabla ^2+r_{\bot }\right) {\bf s}_{\bot }+B_3
{\bf s}_{\bot }^2{\bf s}_{\bot }\right] , 
\end{equation}
where $r_{\bot }$ denotes the corrected mass (see below). Similarly to the
scalar model, this action is critical for an appropriate value of the
disorder, or equivalently of the ``mass''. The recursion relations for 
$r_{\bot }$ and $u_{\bot }\equiv R\left[ \ln b/\left( 4\pi \right) ^3\right]
B_3$ can be read off from Eq.~(\ref{recursion}) by setting $A_i=\bar
A_i=B_2=\bar B_3=0$, as 
\begin{equation}
\label{transrecursion}\left\{ 
\begin{array}{lll}
r_{\bot }^{\prime } & = & b^2\left\{ r_{\bot }+\left[ \left( n-1\right)
+2\right] \left( I+2r_{\bot }\right) u_{\bot }\right\} \\ 
u_{\bot }^{\prime } & = & b^\epsilon \left\{ u_{\bot }+2\left[ \left(
n-1\right) +8\right] u_{\bot }^2\right\} 
\end{array}
\right. \text{.} 
\end{equation}
From these recursion relations we can obtain the static exponent $\nu $
with which the correlation length $\xi $ diverges. By definition $\xi \sim \left(
R-R_c\right) ^{-\nu }\sim \left| r-r_c\right| ^{-\nu }$, which along with 
$\xi^{\prime }=\xi /b$ and $\left( \partial r^{\prime }/\partial r\right) _{
\text{fixed point}}=b^{2-\frac{\left( n-1\right) +2}{\left( n-1\right) +8}
\epsilon }\equiv b^{y_r}$, yields $\left( b^{y_r}\left| r-r_c\right| \right)
^{-\nu }=\left| r-r_c\right| ^{-\nu }b^{-1}$, {\it i.e.}  $\nu y_r=1$, and
\begin{equation}
\label{exponent}\nu =\frac 12+\frac 12\frac{\left( n-1\right) +2}{\left(
n-1\right) +8}\epsilon \text{.} 
\end{equation}
Note that the action of Eq.~(\ref{transaction}) is identical to the critical
action for $n-1$ (weakly coupled) scalar fields, with rotational symmetry.
It follows immediately from this consideration that the recursion relations
and exponents for the longitudinal (Ising-like) fixed point are identical to
those calculated here, with $n-1=1$.

In the reduced action, the transverse components' bare mass becomes 
\begin{equation}
r_{\bot }=B_1+B_2\lambda +\bar B_3\lambda ^2\text{.} 
\end{equation}
The critical line $r^*_{\bot }\left(u_{\bot }\right)$ is in the lower half ($r_{\bot }<0$) 
of the ($r_{\bot},u_{\bot }$) plane. Since $A_0\rightarrow \pm \infty $ as $H\rightarrow \pm
\infty $, Eq.~(\ref{lambdaeq}) implies that $\lambda$ also goes from $-\infty$
to $+\infty$ as the field is increased (similarly to $\tilde A_3<0$
[Eq.~(\ref{renoraction})], we have $\bar B_3<0$). Thus, $r_{\bot }$ crosses
the critical line for a given value of $H$, provided that $B^2_2-4\bar
B_3\left(B_1-r^*_{\bot }\left(u_{\bot }\right)\right)\geq 0$, {\it i.e.}
unless $B_1$ is too negative (for a Mexican hat potential,
we expect $B_1>0$). This may not be true if $A_0$ is a discontinuous
function of $H$. But $A_0\left( H\right) $ is clearly continuous for
critical and higher disorders. (A very high disorder, though, may suppress
this transverse criticality by renormalizing $r_{\bot }$ into large negative
values.) Whether or not the fixed point controlling the transverse
criticality is reachable experimentally depends on what region of the space
of the theory's parameters is swept when the physical quantities at hand in
the experiment are varied. Unlike many equilibrium problems in which the
symmetry is broken by an infinitesimal field, criticality can occur here at
large values of $H$, allowing for non-negligible higher order terms, such as 
$H^2s_{\Vert }^2$ or $H^2s_{\Vert }^2{\bf s}^2$, in the Hamiltonian. These
terms appear in the theory as modified (strong) functional dependences of
the coefficients $A_i$, $\bar A_i$, $B_i$, $\bar B_i$, on the magnetic
field, which, depending on the trend, may favor a transverse instability.

The transverse critical point corresponds to an infinite susceptibility at
some $H_{\bot }$, resulting in a spontaneous {\it transverse magnetization}.
A similar phenomenon was noted for the case of a pure (thermal) system
in Ref.~\cite{dhar}. There, however, the appearance of a transverse
magnetization is due to a magnetic field which oscillates at high frequency.
It is a purely dynamical effect, not observed in the $\eta \rightarrow 0$
limit. In our case, the effect is due to the presence of a quenched
randomness, and its non-equilibrium nature lies in the metastability of
the minima occupied by the system during its history.
This transverse instability, though, differs from the
longitudinal criticality in that it occurs for a range of disorders, rather
than at a specifically tuned amount of randomness. Its underlying physical
mechanism may for example be illustrated by two beads, attached to each
other by a spring, and flowing on the two sides of a Mexican hat's bump, as
it is progressively tilted. At some point, it might become favorable for one
of the beads to jump on the opposite side of the rim, and to continue its
motion next to the other bead, corresponding to a transverse ordering.

At $H_{\bot }$, the transverse components choose one of the many minima,
which breaks the rotational symmetry in the $\left( n-1\right) $-dimensional
transverse space. We therefore have to correct the action of 
Eq.~(\ref{expanded}) for $H>H_{\bot }$, as we did for the longitudinal field. This
however does not alter the analysis of the longitudinal criticality. By
shifting ${\bf s}_{\bot }$, we eliminate the transverse linear term and can
then follow the same procedure as before. We used the fact that the
correction to Eq.~(\ref{expanded}) vanishes for $H\rightarrow \pm \infty $,
and this clearly still holds here.

\subsubsection{${\cal O}\left( n\right) $ criticality}

In the above, we examined two distinct fixed points, corresponding to an
infinite longitudinal susceptibility and a transverse ordering,
respectively. For a specific choice of the theory's parameters, the two
should merge into a single, rotationally invariant (in the $n$-dimensional
space of the fields) fixed point, at which all components become
simultaneously critical. In addition to the magnetic field and disorder, how
many quantities should we tune to reach such an ${\cal O}\left( n\right) $ 
{\it fixed point}? While Eq.~(\ref{expanded}) is a suitable formulation of
the model when the symmetry is broken in the longitudinal direction, the
form of Eq.~(\ref{compact}) is more appropriate close to a fully
rotationally invariant theory. An RG transformation can be carried out for
it, as was done for the action of Eq.~(\ref{expanded}). Since $\sigma \left(
t\right) $ is spatially uniform, only its ${\bf q}=0$ mode is non-vanishing.
The parameter $\sigma $, therefore, does not participate in the coarse
graining transformation and its normalization is given by the rescalings of
the coordinates and the fields, as
\begin{equation}
\sigma ^{\prime }\left( t\right) =\zeta ^{-1}\sigma \left( b^2t\right)
=b^{\frac d2-2}\sigma \left( b^2t\right) . 
\end{equation}
The recursion relations for $H$, $c_1$, and $c_2$, on the
other hand, are obviously given by Eq.~(\ref{recursion}), with $A_0=H$, 
$A_1=B_1=c_1$, $A_3=\bar A_3=B_3=\bar B_3=c_2$, 
and $A_2=\bar A_2=B_2=0$.

By tuning both the magnetic field $H$, and $H_0$, defined as the zero of 
$\sigma $ ($\sigma \left( H_0\right) =0$), to zero, the action of 
Eq.~(\ref{compact}) reduces to the form of Eq.~(\ref{transaction}), with $n$ fields 
${\bf s}$ rather than the $n-1$ components of ${\bf s}_{\bot }$, which is
critical for a given value of the disorder. Consequently, the recursion
relations and exponents are identical to those characterizing the transverse
critical point, but with $n$ replacing $n-1$. Furthermore, since $H=0$ and
the action is fully rotationally invariant, hysteresis is suppressed at the 
${\cal O}\left( n\right) $ critical point; in particular, ${\bf m}\left(
H=0\right) =0$. Finally, by identifying the relevant parameter $\sigma $, we
have established that only a single additional
physical quantity needs to be tuned to reach
the ${\cal O}\left( n\right) $ fixed point (provided that $H_0$ crosses zero
as the physical quantity in question is varied, at $H=0$).

The symmetric fixed point is characterized by three relevant parameters, 
namely $r$ and $H$, with the usual exponents $y_r=2-\epsilon\left(n+2\right)/
\left(n+8\right)$ and $y_H=3-\epsilon/2$, and $\sigma$ with $y_\sigma=d/2-2$,
to ${\cal O}\left( \epsilon\right)$. The parameter $\sigma$ is a measure of
the deviation from a symmetric (non-hysteretic) theory at $H=0$, and,
according to the above exponents, is much less relevant than the other
symmetry breaking term $H$, or of the mass $r$,
to ${\cal O}\left( \epsilon\right)$. Ref.~\cite{leshouches}
notes the fact that the ``critical region'' is
unusually large in the scalar model, as signaled by power laws with
surprisingly high cutoffs even a few percent to one hundred percent away
from the critical disorder, and briefly discusses possible origins of this
phenomenon. This observation, along with the weakness of $\sigma $'s
relevance, suggests in our case that even away from the ${\cal O}\left(
n\right) $ fixed point, an ${\cal O}\left( n\right) $-like behavior might be
displayed at short ranges by the system, before crossing over to an
Ising-like behavior at long enough time and length scales\cite{thermal}.

As mentioned in the introduction, a vanishing magnetization does not a
priori ensure a fully rotationally invariant system; higher moments
could in general display anisotropy. In that case, a complete theory
would break the rotational symmetry, leading to anisotropic exponents
different from $y_r$, $y_H$, and $y_\sigma$ above, which are the outcome
of a restricted action that properly describes only quantities linear in
the spin field. In general we may ask the following question, which
applies to the three cases, Ising, transverse, and ${\cal O}\left(
n\right)$. Would a complete theory yield the same exponents as our
resticted theory? The latter correctly describes the magnetization 
${\bf m}_-\left( t\right)$. That is, if one were able to calculate the
path integral
\begin{equation}
\int {\cal D}\hat s\,{\cal D}s\,{\bf s}\left( {\bf x},t\right) e^{S\left[ \hat
{\bf s},{\bf s+\sigma} \right] }
\text{,} 
\end{equation}
one would obtain ${\bf m}_-$ as a function of ${\bf H}\left( t\right)$,
$R$, and an additional tuning parameter (corresponding to $H_0$), and
thus the associated exponents describing the critical singularity of the
magnetization. It is easy to see that, as for the non-random equilibrium
thermal Landau-Ginzburg model, the exponents $y_r$, $y_H$, and $y_\sigma$
have a one-to-one correspondance with the exponents that characterize
the singular behavior of the magnetization, as well as with those
associated with a number of other quantities, such as the correlation
length. This immediately implies an affirmative answer to the above
question; the requirement that the action should describe the physical
magnetization is a strong enough constraint for the theory to correctly
generate the singularity of, {\it e.g.}, the correlation length.
Evidently, this argument does not apply to the critical behavior of
higher moments of the spin field distribution, and whether or not the 
latter are isotropic at criticality is an interesting open question.
In fact, even the isotropy of $y_r$ at the ${\cal O}\left(n\right)$ 
critical point may seem surprising at first, since the history of the
system apparently sets up an anisotropic context for the spins' motion.
It is consistent, however, with the fact noted in Sec.~IV.A,
that different time slices of the action decouple under the RG in the
$\eta \rightarrow 0$ limit, in which the static exponents are
calculated.

As the reader may have noticed, our recursion relations and exponents in 
$d=6-\epsilon $ are none other than the recursion relations and exponents for
the pure (equilibrium) ${\cal O}\left( n\right) $ model in $d=4-\epsilon $.
Indeed, dimensional reduction to two lower dimensions holds
perturbatively \cite{prb,thesis,dim-red,parisi}. This allows us to
obtain, with no further calculational effort, the expansion for $\nu $ to
higher orders in $\epsilon $ \cite{amit}.

The dynamic exponent $z$, on the other hand, cannot be obtained from a
time-independent model. How is the renormalization procedure modified when 
$\eta \neq 0$? In the coarse-graining process, vertices which couple
different times are generated. In particular, a non-local quadratic term of
the type 
\begin{equation}
\int_{-\infty }^{t^{\prime }}dt\,\hat s_{\Vert }\left( t^{\prime }\right)
s_{\Vert }\left( t\right) f\left( t^{\prime }-t\right) \text{,} 
\end{equation}
is generated in the action, where $f$ is some function (containing free
propagators, disorder, etc). Expanding $s_{\Vert }\left( t\right) $ about 
$t^{\prime }$, this is rewritten as
\begin{equation}
\int_{-\infty }^{t^{\prime }}dt\,\hat s_{\Vert }\left( t^{\prime }\right)
\left\{ s_{\Vert }\left( t^{\prime }\right) +\left. \frac{\partial s_{\Vert
} }{\partial t}\right| _{t^{\prime }}\left( t-t^{\prime }\right)
+...\right\} f\left( t^{\prime }-t\right) \text{.} 
\end{equation}
The first term contributes to the coarse-grained ``mass'', the second
results in a correction to $z$. In fact, the perturbative calculation we
just briefly described is similar to that done by by Krey \cite{krey} 
for the time-dependent thermal random field model, where he calculates $z$ to
${\cal O}\left( \epsilon ^3\right) $, giving to lowest non-trivial order in $\epsilon $,
\begin{equation}
z=2+\frac{n+2}{\left( n+8\right) ^2}\,\epsilon ^2\text{.} 
\end{equation}
(Here, $n$ is the number of components that become massless at criticality.)

\section{Conclusion}

The main aim of the present paper was to extend earlier work on critical
hysteresis to the case of an ordering of continuous symmetry. Our central
result is that, generically, criticality is Ising-like, {\it i.e.} the
critical exponents calculated for a scalar field model \cite{dahmen,prb,thesis}
still hold for a vector field. By tuning a single additional
quantity, however, a fully symmetric fixed point can be reached, to
which ${\cal O}\left( n\right)$-like exponents are associated.
Furthermore, the possibility of a
spontaneous non-equilibrium transverse magnetization is unveiled and
examined using a perturbative renormalization scheme. In addition,
we have clarified several issues pertaining to the path integral formalism,
in particular to the structure of time dependences, and to the problem of
multiple solutions. These analyses are useful for our treatment of the
problem, as well as for a clearer understanding of earlier works.

An interesting question which remains open is that of the lower critical
dimension. In the fully rotational case, naive dimensional reduction
suggests that the lower critical dimension is 4. If this were the case, no
vector criticality should be observed in three dimensions. Several scenarios
are possible. For example, the hysteresis curve may display a jump for low
disorder and be continuous for high disorder, but without the intermediate
limit case of a continuous curve with a diverging slope at a given point. A
more probable scenario is one in which the hysteresis curve is already
smooth for an infinitesimal amount of disorder.

\section*{Acknowledgments}

RdS is grateful to Prof. S. Coleman for illuminating and interesting
discussions, and has benefitted from conversations with Dr. K. Dahmen and
Dr. D. Ertas. This work was supported by the NSF through grant No.
DMR-93-03667.

\appendix 

\section{Expansion about Mean Field Theory}

In Refs.~\cite{dahmen,prb,thesis}, the effective action is written as an
expansion about mean field theory (MFT). In such a formulation, the
linear term in the action (corresponding to $A_0$ in
Eq.~(\ref{expanded})) is vanishing. The ``masses'', {\it i.e.} the
coefficients $r_{\Vert }$ and $r_{\perp }$ (corresponding to $A_1$ and 
$B_1$ in Eq.~(\ref{expanded})) of the parallel and perpendicular
quadratic terms are calculated as the elements of a particular response tensor
within MFT \cite{fisher,dahmen,prb,thesis}. In this appendix, we
calculate $r_{\Vert }$ and $r_{\perp }$, starting from the mean field
equations of motion, and find that, generically, $r_{\Vert }\neq r_{\perp
}$. This appendix thus reaffirms that criticality is Ising-like in
general, as found in Sec.~IV, and makes contact with the earlier
methodology of Refs.~\cite{dahmen,prb,thesis,fisher}.

MFT is defined by an infinite-range coupling, leading to the equations of
motion 
\begin{equation}
\eta \partial _t{\bf s}\left( {\bf x}\right) ={\bf m}\left( t\right) +{\bf H}
\left( t\right) +{\bf f}\left( t\right) {\bf +h\left( x\right)}+c_1{\bf 
s\left( x\right)+}c_2{\bf s\left( x\right) }^2{\bf s\left( x\right)
}\text{,}
\end{equation}
where ${\bf m}\left( t\right) $ is defined self-consistently by ${\bf m}
\left( t\right) =\overline{\left.{\bf s}\left( {\bf x},t\right)\right|_{
{\bf f}=0} }$, and ${\bf f}
\left( t\right) $ is a test field. The ``masses'' are the static components
of the tensor $\delta \bar{\bf s}/\delta {\bf f}-I$, where $I$ is the identity
matrix \cite{fisher,dahmen,prb,thesis}. We have 
\begin{equation}
\label{susceptibility}\frac{\delta {\bf m}}{\delta {\bf H}}=\left. \left( 
\begin{array}{ccc}
\frac{\delta m_1\left( H+H_1,H_2,\cdots \right) }{\delta H_1} & \frac{\delta
m_1\left( H+H_1,H_2,\cdots \right) }{\delta H_2} & \cdots  \\ 
\frac{\delta m_2\left( H+H_1,H_2,\cdots \right) }{\delta H_1} & \frac{\delta
m_2\left( H+H_1,H_2,\cdots \right) }{\delta H_2} & \cdots  \\ 
\vdots  & \vdots  & \ddots 
\end{array}
\right) \right| _{H_1=H_2=\cdots =0}\equiv \left( 
\begin{array}{cccc}
\chi _{\Vert }\left( H\right)  & 0 & 0 & \cdots  \\ 
0 & \chi _{\bot }\left( H\right)  & 0 & \cdots  \\ 
0 & 0 & \chi _{\bot }\left( H\right)  & \cdots  \\ 
\vdots  & \vdots  & \vdots  & \ddots 
\end{array}
\right) \text{.}
\end{equation}
(We consider $\eta \rightarrow 0$, so responses are non-vanishing only at
equal times, and only the static part remains. Also, in MFT, the
magnetization is always parallel to the magnetic field, leading to vanishing
off-diagonal elements and to the relation $\chi _{\bot }=m/H$ \cite{ertas}.)
In order to calculate the relevant tensor, let us split the field ${\bf f}$ 
between ${\bf m}$ and ${\bf H}$, in such a way as to satisfy 
Eq.~(\ref{susceptibility}) for effective magnetization and magnetic
field. That is, we write the equations of
motion in the following fashion, with corrected ${\bf m}$ and ${\bf H}$, and
with no additional field, 
\begin{equation}
\eta \partial _ts_{\Vert }=\left( m+\frac{\chi _{\Vert }}{1+\chi _{\Vert }}%
f_{\Vert }\right) +\left( H+\frac 1{1+\chi _{\Vert }}f_{\Vert }\right)
+h_{\Vert }+c_1s_{\Vert }+c_2{\bf s}^2s_{\Vert }
\end{equation}
for the longitudinal component, and 
\begin{equation}
\eta \partial _t{\bf s}_{\bot }=\left( \frac{\chi _{\bot }}{1+\chi _{\bot }}%
{\bf f}_{\bot }\right) +\left( \frac 1{1+\chi _{\bot }}{\bf f}_{\bot
}\right) +{\bf h}_{\bot }+c_1{\bf s}_{\bot }+c_2{\bf s}^2{\bf s}_{\bot }
\end{equation}
for the transverse components. Whence
\begin{equation}
\frac{\delta \bar{\bf s}}{\delta {\bf f}}=\left( 
\begin{array}{cccc}
\frac 1{1+\chi _{\Vert }^{-1}} & 0 & 0 & \cdots  \\ 
0 & \frac 1{1+\chi _{\bot }^{-1}} & 0 & \cdots  \\ 
0 & 0 & \frac 1{1+\chi _{\bot }^{-1}} & \cdots  \\ 
\vdots  & \vdots  & \vdots  & \ddots 
\end{array}
\right) \text{,}
\end{equation}
and 
\begin{equation}
r_{\Vert }=-\frac 1{1+\chi _{\Vert }} \text{,~and~}r_{\bot }=-\frac 1{1+\chi
_{\bot }}=-\frac 1{1+m/H}.
\end{equation}
Thus, $r_{\Vert }\left( H\right) \neq r_{\bot }\left( H\right) $ in general.


\begin{thebibliography}{99}
\bibitem{precession}  To be more accurate, the spins not only follow their
local minima, but also precess. However, since $H$ can be increased
arbitrarily slowly, the precession time scale can be separated as much as we
want from the ``$\Omega ^{-1}$'' time scale. Then, considering the motion on
an intermediate time scale, we find that the precession averages out, and we
are effectively left with the above Langevin equation.
\bibitem{jacobian}  The path integral is defined as the limit of a
time-discretized formulation. In the simplest (Ito) discretization scheme in
which the response to a force at time $t$ occurs at time $t+\epsilon $, the
Jacobian is clearly independent of the spin field $s$. In this case, the
contraction of $\hat s$ and $s$ (response function) vanishes at equal time.
Other discretization procedures lead to more complicated Jacobians, which,
in all cases, cancel the naive equal time responses resulting from the
discrete equations. (See also \cite{fisher} and references therein.)
\end{thebibliography}
\end{document}